# Survey of Key Distribution Schemes for Wireless Sensor Networks

Pratik P. Chaphekar
Oklahoma State University

***Abstract.*** This paper presents a survey of key distribution schemes for wireless sensor networks. This survey includes the new approach of key distribution using the piggy bank method. Different Network architectures and different key pre-distribution schemes are described. The survey includes the use of the piggy bank approach to cryptography in which part of the key is pre-distributed and the remainder is varied in the application.

## 1. Introduction

A wireless sensor networks (WSNs) consist of large number of sensor nodes that has limited power resources, computation and communication distributed in a designed area without any fixed structure. Wireless sensor networks are used in applications involving high security such as border security, military target tracking and scientific research which are in dangerous environment.

WSN security has six challenges (i) To perform communication in the wireless mode, (ii) Operate lacking fixed infrastructure, (iii) Sensor nodes have limited resources (iv) WSN can be huge and dense, (v) Prior to deployment the network topology is unknown, (vi) Risk of physical attacks on unattended sensors are high. Thus, security is a primary concern especially in hostile environment like military areas. For instance, an adversary could capture sensor nodes, intercept transmitted messages, and propagate fake messages to the networks. Researchers have investigated the security of WSNs from many different perspectives [1]-[15]. Key distribution plays an important role for security in wireless sensor networks. Sensor node needs to adapt to the environment and establish a secure network by using techniques like Key pre-distribution and exchanging information with their immediate neighbor's or computationally strong nodes.

Based on application some of the solutions for wireless network security can be evaluated as follows: (i) Different network architecture like distributed or hierarchical, (ii) Key management such as pre-distributed or dynamically generated pairwise, group-wise or network wise keys, (iii) Different styles of communication like pair-wise(unicast), group-wise(multicast) or network-wise(broadcast), (iv) Security requirements like authentication, integrity and confidentiality.

A sensor node is a small device that consists of processing unit, sensing unit, transceiver and power unit as shown below.

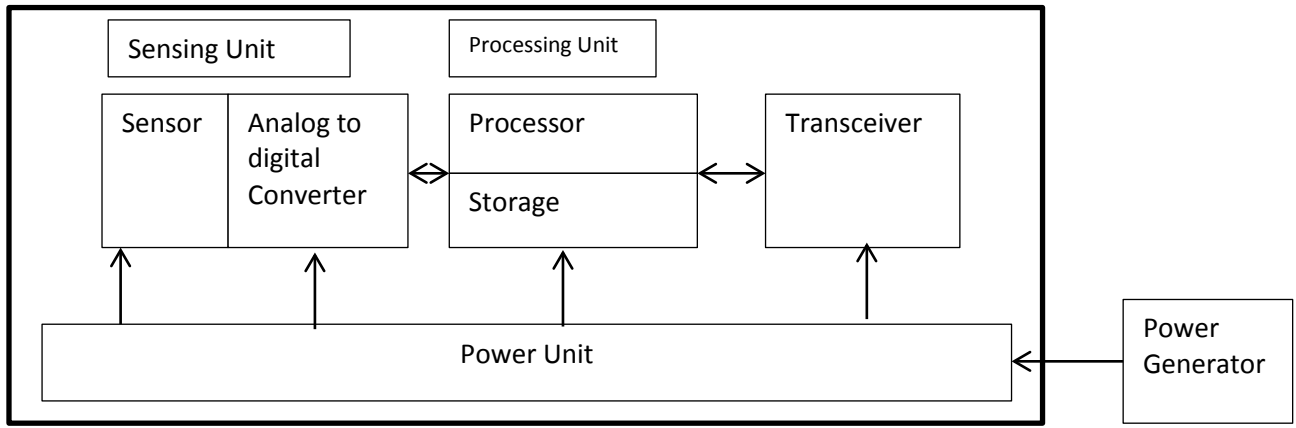

Fig 1: Components of a sensor node [7]

An example of a sensor node hardware configuration is the Berkeley Mica Motes which features an 8-bit 4 Mhz Atmel ATmega 128L processor with 128Kbytes program store and 4K bytes SRAM [7]. The processor allows limited set of RISC-like instruction set, without any support to rotate or multiplication or variable-length shifts. The peak rate of ISM band radio receiver communication is of 40Kbps at a range of up to 100 feet.

We will focus predominantly on key management techniques but we stress that the question of the randomness of the key sequences used is an important issue [16]-[25] as is also the question of efficient matrix computations [26]-[29]. Other issues include those relevant in specific applications such as health care networks [30], the problem of security attacks [31], anomaly detection [32]. WSN's face multiple threats and these include: communication attack; denial of service attack; node compromise; impersonation attack; and protocol-specific attack.

As the nodes have limited resources; key management in WSNs is a problem. In the literature, key management protocols are based on either symmetric or asymmetric cryptographic functions. Key management protocols based on public key cryptographic (asymmetric functions) are not appropriate due to resource limitations in sensor nodes, thus key pre-distribution a particular symmetric approach is deployed in WSNs, which reduces the cost of key establishment. However, it appears that the piggy bank version of public key cryptography [33] can be adapted for sensor networks by pre distributing elements of the key.

## 2. Network architectures

Before we start with key pre-distribution scheme let us understand in brief the two principal network architecture that is used. Typically, a WSN grows in an ad hoc manner and therefore it is unstructured. Other networks are deployed in a pre-planned manner and categorized as structured. Networks may also be categorized as having hierarchical or distributed architectures.

1) *Hierarchical WSN (HWSN) [8]:*

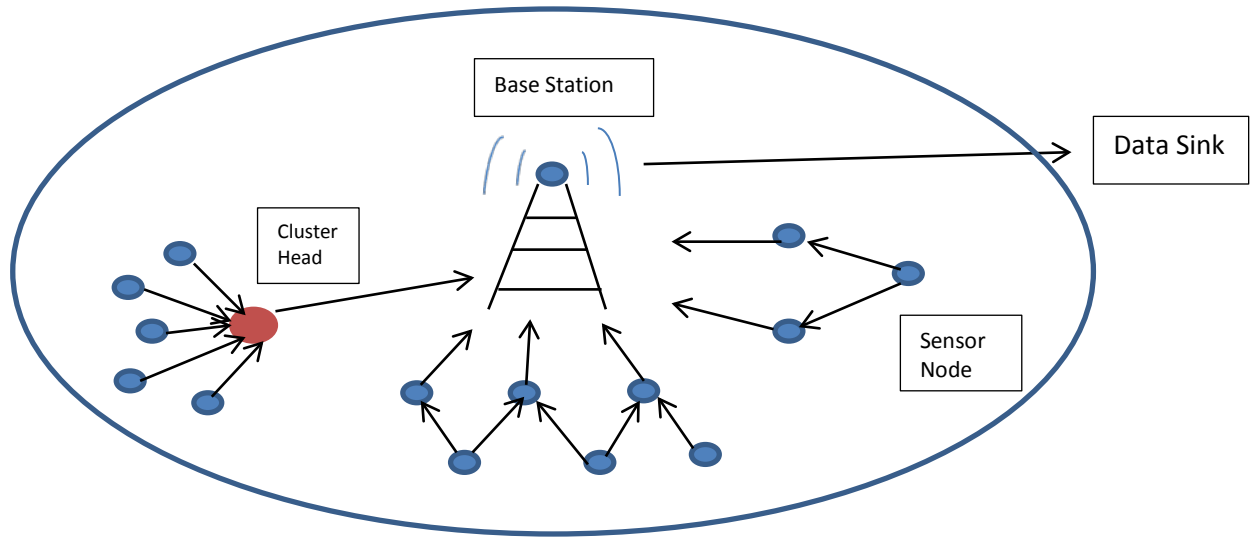

Fig 2: Hierarchical WSN [8]

The figure above shows the Hierarchical WSN, nodes are arranged in hierarchy based on their capabilities. As you can see from the figure above there are 3 types of nodes: Base station, cluster head and sensor nodes, base station is powerful than the other two nodes. A base station is the heart of a WSN, it is a powerful data processing /storage center, or an access point to human interface, it is typically a gateway to another network. Apart from this, it also performs tasks like collecting sensor readings, performing costly operations on behalf of sensor nodes and managing the network. In some applications it is used as key distribution center has it is assumed to be the most trusted and tamper resistant.

Transmission power of a base station is sufficient enough to reach to all sensor nodes; however sensor nodes depend on ad hoc communication to reach to base station. Sensor nodes form a dense network where a cluster of sensors residing in a particular area may provide close or similar readings. Sensor nodes are deployed to base station by one or more hop neighborhood. Cluster head are the nodes with better resources and may be used to collect and merge the local traffic and send it to the base stations. The data flow in a network like this can be pair-wise (unicast) among sensor nodes, group-wise (multi-cast) within a cluster of sensor nodes, and network-wise (broadcast) from base stations to sensor nodes.

The table below shows the solution to key distribution problem:

| Problem | Keying Style |
|---|---|
| Pairwise | BS oriented |
| | Master key |
| Group-wise | Asymmetric keys |
| | Symmetric keys |
| Network-wise | Master key |
| | TESLA based |

Table 1: Solutions on key distribution problem in Hierarchical WSN (TESLA: Timed Efficient Stream Loss-tolerant Authentication.)

*2)     Distributed WSN (DWSN) [8]:*

In DWSN the sensor nodes are scattered randomly all over the target area as shown in figure below. Thus DWSN lack fix infrastructure and network topology is not known prior to deployment. On deployment, sensor nodes figure out its neighbor by scanning its radio coverage area. The flow of data is similar to that in HWSN with a difference that network-wise (broadcast) can be sent by every sensor nodes. In this, sensor nodes either use keying materials to dynamically generate pairwise and group-wise keys or use pre-distributed keys directly.

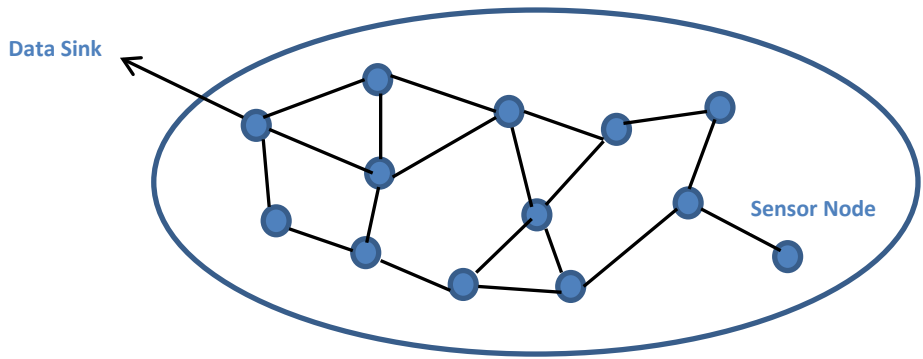

Fig 3: Distributed WSN [8]

To overcome a difficulty of an efficient way of distributing keys and keying materials to sensor nodes prior to deployment is to use one of the three approaches (i) Probabilistic (ii) Deterministic (iii) Hybrid.

In probabilistic approach, there exist a key pool and key chains are randomly selected from key-pool and distributed to sensor nodes, whereas in deterministic approach the key-pool and key-chains are designed using various deterministic processes to provide better key connectivity. In Hybrid approach, probabilistic approach is used on deterministic approach to improve the scalability and resilience.

The table below shows the solution to pair-wise and group-wise key distribution

| Problem | Approach | Mechanism | Keying style |
|---|---|---|---|
| Pair-wise | Probabilistic | Pre-distribution | Random Key-chain |
| | | | Pairwise key |
| | Deterministic | Pre-distribution | Pairwise key |
| | | | Combinatorial |
| | | Dynamic Key Generation | Master Key |
| | | | Key Matrix |
| | | | Polynomial |
| | Hybrid | Pre-distribution | Combinatorial |
| | | Dynamic Key Generation | Key Matrix |
| | | | Polynomial |
| Group-wise | Deterministic | Dynamic Key Generation | Polynomial |

Table 2: Solutions on key distribution problem in Distributed WSN

Now we discuss various schemes proposed for key pre-distribution in WSNs

## 3. Polynomial based key pre-distribution

We discuss how to pre distribute a pairwise keys.

The (key) setup server randomly generates a bivariate t-degree polynomial $f(x,y) = \sum_{i,j=0}^{t} a_{ij} x^i y^j$ over a finite field $F_q$, where q is a prime number that is large enough to accommodate a cryptographic key, such that it has the property of f (x, y) = f (y, x). (We assume all the bivariate polynomials have this property without explicit statement.) Each sensor is associated to a unique ID in this system. For each sensor 'i', the setup server computes a polynomial share of f(x, y), that is, f(i, y ). For any two sensor nodes i and j, node i can compute the common key f(i, j ) by evaluating f (i, y ) at point j, and node j can compute the same key   f (j, i) = f (i, j ) by evaluating  f(j, y) at point i. [1]

## 4. Polynomial pool based key pre-distribution

As the name suggest, in this technique we have a pool of randomly generated bivariate polynomial; the method used for generating this bivariate polynomial is based on the polynomial based key pre-distribution as discussed above. There are two cases of the polynomial pool. When the polynomial pool has only one polynomial, the general framework degenerates into the polynomial-based key pre distribution. When all the polynomials are 0-degree ones, the polynomial pool degenerates into a key pool Random Subset assignment  key pre-distribution.[2][3]

## 5. Grid based key pre-distribution

Assume a sensor network with at most N sensor nodes. The grid based key pre distribution scheme then constructs a m×m grid with a set of 2m polynomials        { $f_i^c(x,y)$ , $f_j^r(x,y)$} $_{i=0,...,m-1}$},where m = [√N ].As shown in Figure below , each row j in the grid is associated with a polynomial $f_j^r(x,y)$, and each column i is associated with a polynomial $f_i^c(x,y)$. The setup server assigns each sensor in the network to a unique intersection in this grid. For the sensor at the coordinate (i, j ), the setup server distributes the polynomial shares of $f_i^c(x,y)$ and $f_j^r(x,y)$ to the sensor. As a result, sensor nodes can perform share discovery and path discovery based on this information and eventually can pre-distribute the keys.[4]

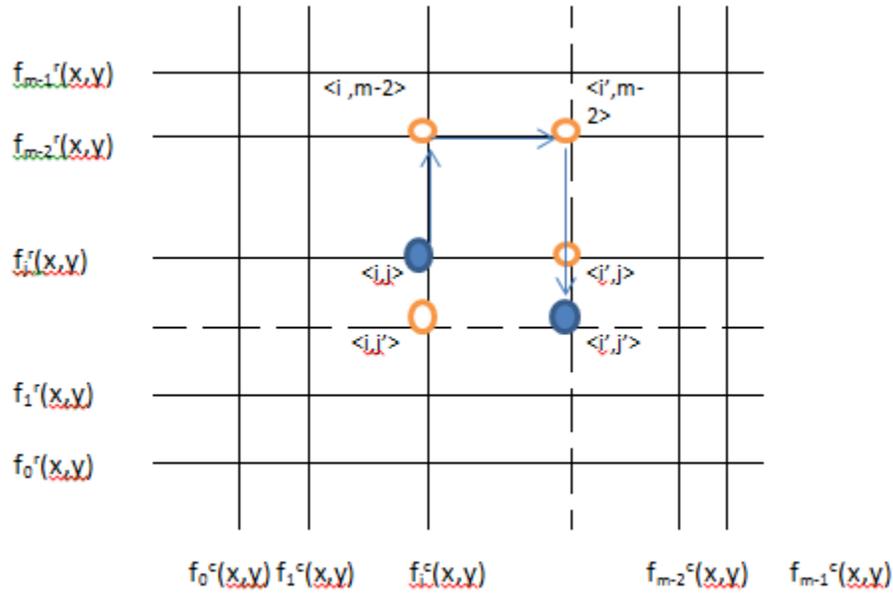

Figure 4: The grid [4]

## 6. Hyper-cube multivariate Scheme (HMS)

It is basically a Threshold based scheme, a number of multivariate polynomial is assigned to every point on the hypercube, and a hypercube is designed in the multidimensional space. The points on the hypercube are uniquely assigned to the sensors. Using this technique, a direct key is established between any two sensors at Hamming distance of one from each other. Other sensors are also able to establish indirect keys.[5]

## 7. Key Pre-distribution scheme using Combinatorial design

This scheme is a deterministic key pre-distribution in WSN. First to understand the design we need to define some terms [10]

**Definition 1: Balanced Incomplete Block Design (BIBD)**: A set system or design is a pair (X,A), where A is a set of subsets of X, called blocks. The elements of X are called varieties. A BIBD $(v,b,r,k,\lambda)$ is a design which satisfy the following conditions:

  i.    |X| = v. |A| = b,
  ii.   Each subset in A contains exactly k elements
  iii.  Each variety in X occurs in r many blocks
  iv.   Each pair of varities in X is contained in exactly λ blocks in A.

A BIBD(v,b,r,k,λ) design can be represented by a incidence matrix M = [$m_{ij}$] of dimension v x b with entries 0 and 1, $m_{ij}$ = 1, if the ith variety is present in the jth block and 0 otherwise.

Definition 2 : suppose that (X,A) is a system, where

$$X = \{x_i : 1 \leq i \leq v\}$$

And

$$A = \{A_j : 1 \leq j \leq b\}$$

The dual set system of (X,A) is any set isomorphic to the set system (X',A') where

$$X' = \{x'_j : 1 \leq j \leq b\}.$$
$$A' = \{A'_i : 1 \leq i \leq v\}.$$

And where

$$x'_j \in A'_i \iff x_i \in A_j$$

It follows that if we take the dual of a BIBD(v,b,r,k,λ), we arrive at a design containing b varities, v blocks each block containing exactly r varieties and each variety occurring in exactly k blocks. We also note that any two blocks contain λ elements in common.

**Definition 3 Symmetric BIBD or SBIBD:** When v = b , the BIBD is called SBIBD and denoted by SB[v, k;λ].

**Definition 4 Pairwise balance design (PBD):** It is a design in which each pair of points occurs in λ blocks, for some constant λ, called the index of the design.

**Definition 5:** The intersection number between any two blocks is the number of elements common to the blocks.

**Definition 6 :** Let the intersection numbers between any of the blocks in a BIBD be $\mu_1, \mu_2, \ldots \mu_x$. let M = { $\mu_i$ : i = 1,2,....,x } let μ = max{ $\mu_1, \mu_2, \ldots \mu_x$ }. μ is called the linkage of the design.
Note: For a SBIBD, |M|= 1 and , μ = λ.

**Design:** A BIBD (v,b,r,k,λ) can be mapped to a sensor network containing v keys in the key-pool. There are b sensor nodes, each node containing k keys are each key occurring in r nodes. Any pair of keys occurs in λ blocks. Consider two BIBDs : D1 = ($v_1,b_1,r_1,k_1,\lambda_1$) and D2 =($v_2,b_2,r_2,k_2,\lambda_2$). Let $M_1$ = [$m_{ij}'$] and $M_2$ = [$m_{ij}''$] be the respective incidence matrices, therefore the dimensions of $M_1$ and $M_2$ are $v_1$x $b_1$ and $v_2$ x $b_2$ respectively. A requirement of the design is that $k_1=v_2$. The matrix M is constructed in the following way
1. For every column j of $M_1$ replace $m_{ij}'$ by a row of $M_2$ if $m_{ij}'$ = 1. For each i replace $m_{ij}'$ by a different row of $M_2$.
2. For every column j of $M_1$ replace $m_{ij}'$ by a row vector of length $b_2$ containing all zero's if $m_{ij}'$=0.

The result of the following operations is a matrix M of dimension $v_1$ x $b_1b_2$. We call the design D and represent it by D = $D_1 \bowtie D_2$. It is said D is the expanded design of $D_1$ and $D_2$. The blocks in D which arise from a given block in $D_1$ are said to belong to the same group. Now we map this design D to a sensor network consisting of $b_1b_2$ nodes, each node consisting of $k_2$ keys. The key-pool consists of $v_1$ keys [10].

# 8. A Robust Key Pre-distribution protocol for Multi-phase WSN

From the components of sensor node we see that they are battery operated and their life time is much smaller as compared to the lifetime of the whole network and hence we need to deploy new sensor nodes for the disappearing ones in the network to assure good network connectivity. The process of deploying new sensors at regular period is called generations. The time between two successive generations is called generation period. Gw - generation window:- a number which is assumed that a sensor lifetime is bounded of generations. The generation period is assumed to be 1. The time period for a node deployed at generation j to establish a secret channel with any other sensor deployed is |j-k,j+k| where k is a integer and its assume that k = Gw in this scheme. If the value of k is less than Gw the newly deployed node can establish a secure channel with a subset of the network sensors. Each sensor is assigned A two different key rings: $FKR_A$ and $BKR_A$, where a key ring is a subset of key pool namely $FKR_A \subset FKP$ and $BKR_A \subset BKP$. The abbreviation and their full form are listed below

Gw: generation window, n:last generation of the network, A: sensor A, $kr^j_A = (FKR^j_A, BKR^j_A)$ : key ring of A at gen. j, $FKR^j_A$: forward key ring of A at gen. j, $BKR^j_A$: backward key ring of A at gen. j, m:key ring size, $FKP^j$: Forward key pool at gen. j, $BKP^j$: Backward key pool at gen. j, P:key pool size, $fk^j_t \in FKP^j$: t-th f key at gen j, $bk^j_t \in BKP^j$: t-th b key at gen j, h:secure hash function $h:\{0,1\}^* \rightarrow \{0,1\}^{160}$, f:hash function $f:\{0,1\}^* \rightarrow \{0,1\}^{\log_2(P)}$, RKP: key management defined in [2], Rok: Robust key predistribution scheme.

Key pool Generation:
In this scheme the key pool evolve with time and are updated at each generation. Initially the forward key pool contains P/2 random keys. Each key is updated by hashing the current key with secure hash function h, such as SHA1. When the network is deployed the forward key pool is defined as
$$FKP^0 = \{fk^0_1, fk^0_2, ....., fk^0_{P/2}\}$$
$fk^0_i$ is randomly generated. At generation j+1, the fkeys are refreshed as follows
$$FKP^{j+1} = \{fk^{j+1}_1, fk^{j+1}_2, ....., fk^{j+1}_{P/2}\}$$
Where $fk^{j+1}_t = h(fk^j_t)$.
The backward key pool is generated in the similar fashion.

Key ring assignments:
By using a pseudorandom function, subkeys are assigned to each node A deployed at generation j and each node is configured with m/2 subkeys from the forward and backward key pools. The first subkey of the forward key ring will be the subkey with index $f(id_A||1||j)$ of the forward key pool, where f(.) is for example a hash function with range[1;m]. The first backward subkey of backward key ring will be the subkey with index $f(id_A||1||j)$ of the backward key pool and so on.
We can say that a node A is configured with a key ring, $kr^j_A = (FKR^j_A, BKR^j_A)$ defined as follows
$$FKR^j_A = \{fk^j_t | t = f(id_A||1||j), i = 1,2,....,m/2\}$$
$$BKR^j_A = \{bk^{j+Gw-1}_t | t = f(id_A||1||j), i = 1,2,....,m/2\}$$
Where m is the size of the whole key ring.

Depending on the deployment of the owner sensor the key rings are strictly bounded to them, thus the nodes need to be loosely time synchronized.

Updation of a key ring by a node A for the generations of i can only be between j and j+Gw. The lifetime of key ring is limited. Node A cannot compute the bkeys for the generation after generation j+Gw, it cannot update $kr_A$ beyond generation j+Gw, it also cannot recover any fkeys for the generation before j and it cannot compute $kr_A$ for the generation before j. This factor limits the power of the attacker and attacker can use the keys for a limited time span.

Establishing a secure channel:
Neighbor discovery process starts when a node is deployed, consider a node A deployed at generation j, it broadcast a message containing its identifier $id_A$ and its generation, the nodes receiving its message are able to reconstruct the list of the key indexes in $kr_A$ and identify the keys that they have in common. Consider a neighbor node B which constructs the set { t |t = $f(id_A||l||j)$ , l = 1,2,....,m/2} of the key indexes and scan to see if it shares at least one index t with A. If it shares, then node B replies with its identifier $id_B$ and its generation. Thus A identifies the list $t_1$, $t_2$,...., of all the key indexes in common. After this step both the nodes calculate their overlapping generations which is the set of generations in which they can communicate. The overlapping generation will be all the generation in the interval |j,i+Gw| where i is the generation when B was deployed and j when A was deployed and i ≤ j. With the common key indexes $(t_1,t_2,...,t_z)$ between the two nodes they can communicate their forward and backward keys and eventually compute their secret key which is

$K_{AB}$ = $h(fk^j_{t1} || bk^{i+Gw-1}_{t1} || fk^j_{t2} || bk^{i+Gw-1}_{t2} || ....... || fk^j_{tz} || bk^{i+Gw-1}_{tz})$

The key $k_{AB}$ can be used to establish a secure channel between A and B. The forward key should be erased by a node at each new generation, this prevents adversary from compromising forward keys.[9]

## 9. Matrix Based Key Agreement Scheme
The basic algorithm is as follows [6]:

**Pre-deployment:**
A symmetric random matrix K with elements in $Z_p$ is to be chosen.
We need to find two matrix X and Y such that XY=K
Assign nodes with random row - column pairs from X and Y, for example if a node i receive the $r^{th}$ row of X then it also receives the $r^{th}$ column of Y.

**Key Agreement:**
Any two nodes i and j agree on an encryption key by exchanging their columns of Y and compute key $K_{ij}$ = $row_{node(i)}(X).col_{node(j)}(Y)$ = $row_{node(j)}(X).col_{node(i)}(Y)$= $K_{ji}$ , $K_{ij}$ = $K_{ji}$ since K is symmetric. $row_{node(i)}(X)$ refer to row of X that was assigned to node i .

Commuting matrices can also be used; it eliminates the requirement of K being symmetric. The algorithm is as follows

**Pre-deployment:**
Choose two q x q matrices X and Y such that XY = YX and Y is symmetric.
Randomly pick r from a uniform distribution over [1,q].
Assign node i randomly chosen $r^{th}$ row and column of X and the $r^{th}$ column of Y.

Two nodes I and j agree on key as follows
Node i send its column of Y to node j.
Node j send its column of Y to node i.
Node i computes $K_{ij}$= $row_{node(i)}(X).col_{node(j)}(Y)$ and node j computes $K_{ij}$= $col'_{node(i)}(Y).col_{node(j)}(X)$.
Node i computes $K_{ji}$= $col'_{node(j)}(Y).col_{node(i)}(X)$. and node j computes $K_{ji}$= $row_{node(j)}(X).col_{node(i)}(Y)$.
Key used is computed as K=Hash($K_{ij}$|| $K_{ji}$).
Where $col'_{node(i)}(Y)$ is the transpose of column Y assigned to node i.

**Scheme 1:**
Let K be a matrix of size q x q , therefore X and Y are of size q x m and m x q respectively. Number of elements in the upper triangle of matrix including the diagonal is q(q+1)/2 generated from 2qm elements. For each key there are p possibilities which are equally likely.

**Scheme 2:**
In this scheme the commuting matrices X and Y requires to be a square matrix q x q where q ≤ N where N is the number of nodes in a network. K = Hash ($K_{ij}$|| $K_{ji}$) is the common key between nodes i and j. [5]

**Example:** We will show a simple example of Scheme 1 using symmetric matrix.
Let K be a symmetric, K=$\begin{bmatrix} 2 & 4 \\ 4 & 5 \end{bmatrix}$ then we find 2 matrices X and Y such that XY = K
X = $\begin{bmatrix} 2 & 0 \\ 1 & 3 \end{bmatrix}$ Y = $\begin{bmatrix} 1 & 2 \\ 1 & 1 \end{bmatrix}$ , then we can calculate for i = 1 and j = 2, $K_{ij}$ = $row_1(X).col_2(Y)$ = 4 and $K_{ji}$ = $row_2(X).col_1(Y)$ = 4.

## 10.     The piggy bank based methods

The idea of locking and sealing a piggy bank to transport message is extended to data communication [33]. Once a secret message is inserted into the piggy bank and locked, than the message is not accessible until the box is opened with an appropriate key. We first discuss a three-stage protocol in which both parties use lock protected by tamper-proof seal of each parties and then discuss the piggy bank trope. Suppose Alice wants to send a message to Bob, Alice inserts his message in a locked box which is transported to Bob; Bob in turn puts his own lock on the box which is transported back to Alice. Alice unlocks the lock using his key $K_a$ and sends the box again to Bob and then Bob unlocks it with his key $K_b$. By unlocking with their key they ensure that the lock is not tampered.

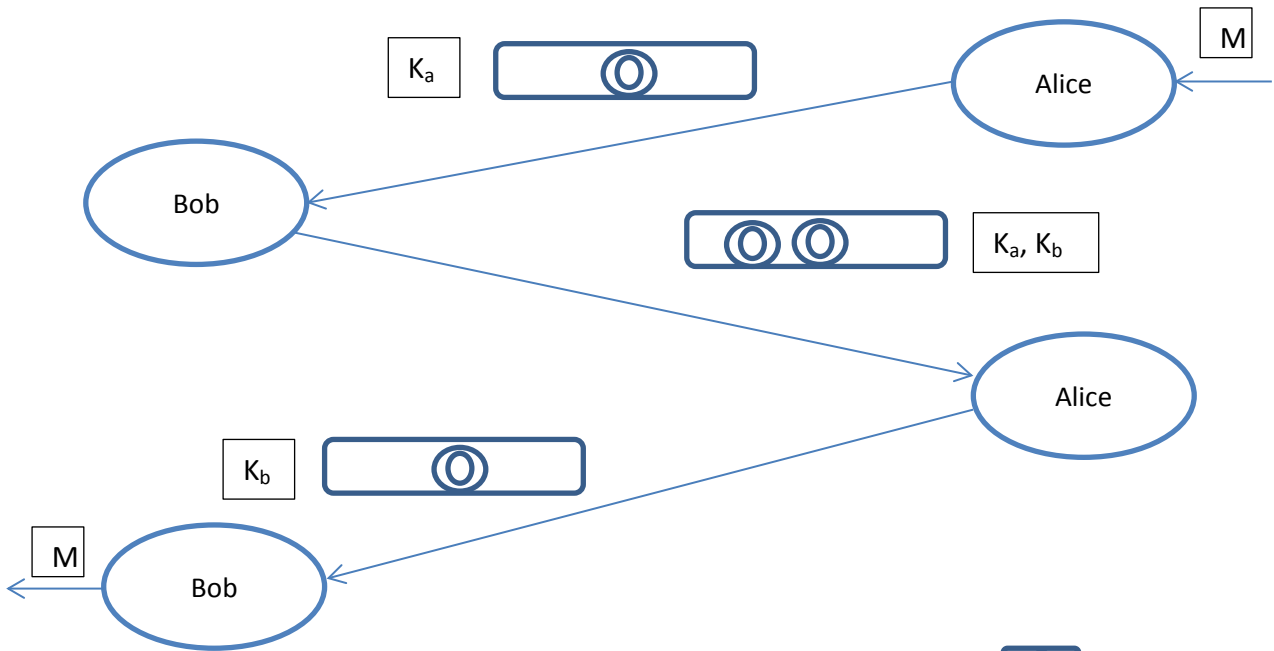

Figure 5: Three-stage protocol using locks 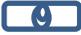.

The Piggy Bank Scheme: The steps involved in this trope are as shown in figure 6

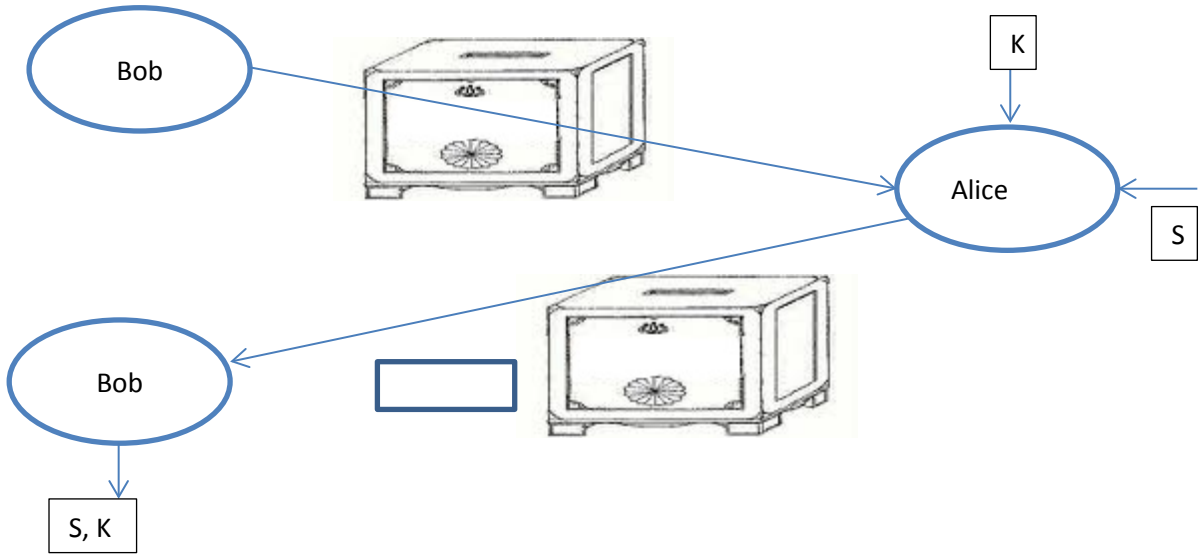

Figure 6: Piggy Bank Cryptographic Trope; the secret letter is represented by

An empty locked piggy bank is sent by Bob to Alice; Alice on receiving the box deposits the secret message along with the decryption key of a coded letter and sends it back to Bob. Alice also sends an encrypted letter to authenticate the contents of the locked piggy bank box. Bob opens the box, obtains the secret and also reads the coded letter. The details of this method are available in [33].

Notice that the communication from Alice to Bob consists of two separate messages. One of these, represented by the box can be pre distributed within the network. This reduces the computation burden on the second transformation which now can be used with reduced complexity in a manner that is appropriate for the reduced computational capacity of the sensor nodes.

In order to understand how it can be implemented in real system let us explain the working with the help of an example. The protocol is implemented in 3 steps:

Step 1: Bob selects a random number say 11 and the piggy bank transformation is represented by a one-way transformation $f(11) = 11^{1037} \mod 323 = \mathbf{7}$ where 323 is a composite number with factors known only to Bob; 1037 is the publicly known encryption exponent. The composite number required for computation is provided by Bob to Alice by some secured channel.
Step 2: Bob sends $f(11) = 7$ to Alice who multiples it with her secret key 3 and sends $f(11)*3 + 17 \mod 323 = \mathbf{38}$ in one communication and $f(3)=3^{1037} \mod 323 = \mathbf{29}$ in another communication.
Step 3: Bob uses his secret inverse transformation function to recover the secret key of Alice; first he recovers 3 and then 17. The inverse transformation is as follows
$29^5 \mod 323 = \mathbf{3}$
$3 * 7 + K = 38$ therefore $K = \mathbf{17}$

Thus we can see from the above calculation Bob retrieves the secret keys of Alice.
The above steps can be shown pictorially as follows

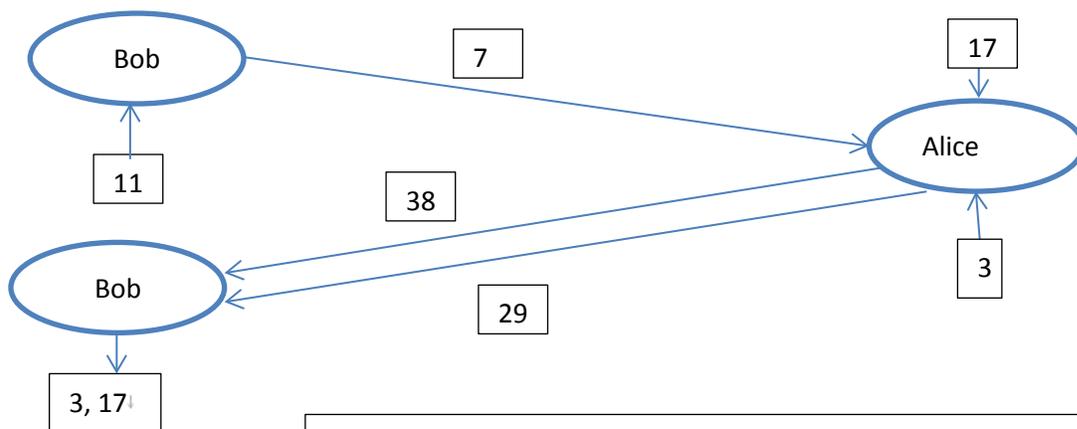

Figure 7: Piggy Bank Cryptographic Trope implementation

Since a sensor node has limited resources, normally public cryptography is not feasible. However, we can use the piggy bank approach in WSN by reducing the computation involved in the process. This can be done by hardwiring one of the secret keys. Once a secret key is hardwired, we can use smaller composite numbers without weakening the level of security.

## 11. Conclusions

This paper presented a survey of key distribution schemes for wireless sensor networks. This survey includes the new approach of key distribution using the piggy bank method. Key management protocols based on public key cryptographic (asymmetric functions) are not appropriate due to resource limitations in sensor nodes, thus key pre-distribution a particular symmetric approach is deployed in WSNs, which reduces the cost of key establishment. However, it appears that the piggy bank version of public key cryptography can be adapted for sensor networks by pre distributing elements of the key.